\documentclass[twocolumn]{aastex631}

\begin{document}

\title{The Impact of Enhanced EUV Flux on the Upper Atmosphere of Earth-like Exoplanets}

\correspondingauthor{Lukas Hanson}
\email{lukas\_hanson@student.uml.edu}

\author[0009-0001-8510-1729]{Lukas Hanson}
\affiliation{Lowell Center for Space Science and Technology, University of Massachusetts Lowell, 600 Suffolk Street, Lowell, MA 01854, USA}

\author[0000-0003-3721-0215]{Ofer Cohen}
\affiliation{Lowell Center for Space Science and Technology, University of Massachusetts Lowell, 600 Suffolk Street, Lowell, MA 01854, USA}

\author[0000-0001-6933-8534]{Aaron J. Ridley}
\affiliation{Climate and Space Sciences and Engineering, University of Michigan, Ann Arbor, MI, USA}

\author[0000-0001-9843-9094]{Alex Glocer}
\affiliation{NASA Goddard Space Flight Center, Greenbelt, MD, USA}



\begin{abstract}

Identifying Earth-like planets outside out solar system is a leading research goal in astronomy, but determining if candidate planets have atmospheres, and more importantly if they can retain atmospheres, is still out of reach. In this paper, we present our study on the impact of enhanced EUV flux on the stability and escape of the upper atmosphere of an Earth-like exoplanet using the Global Ionosphere and Thermosphere Model (GITM). We also investigate the differences between one- and three-dimensional solutions. We use a baseline case of EUV flux experienced at the Earth, and multiplying this flux by a constant factor going up to 50. Our results show a clear evidence of an inflated and elevated ionosphere due to enhanced EUV flux, and they provide a detailed picture of how different heating and cooling rates, as well as the conductivity are changing at each EUV flux level. Our results also demonstrate that one-dimensional solutions are limited in their ability to capture a global atmosphere that are not uniform. We find that a threshold EUV flux level for a stable atmosphere occurs around a factor of 10 times the baseline level, where EUV fluxes above this level indicate a rapidly escaping atmosphere. This threshold EUV flux translates to about 0.3AU for a planet orbiting the Sun. Thus, our findings indicate that an Earth-like exoplanet orbiting its host star in a close-in orbit is likely to lose its atmosphere quickly.

\end{abstract}


\section{Introduction} \label{sec:intro}

Since the first confirmed detection of an exoplanet in the early 1990s \citep{1992Natur.355..145W,1995Natur.378..355M}, more than 5,000 exoplanets have been discovered. Since then, the question whether planets exist outside of our solar system has shifted to the question of whether Earth-like, habitable planets exist outside our solar system and can they be detected. In 2017, seven rocky, Earth-sized planets were discovered around the M-dwarf star TRAPPIST-1 \citep{Gillon_2017}. One of the missions of the James Webb Space Telescope (JWST) is to measure the atmospheric properties of exoplanets, making the TRAPPIST system an excellent target for JWST. Furthermore, future missions will add to our ability to detect and characterize exoplanetary atmospheres, such as PLATO \citep{Rauer_2014}, ARIEL \citep{2016SPIE.9904E..1XT}, JASMINE \citep{kawata2024jasminenearinfraredastrometrytime}, and Habitable Worlds Observatory (HWO) \citep{mamajek2024nasaexoplanetexplorationprogram}. 

In its simplified definition, habitability depends on the planet's ability to sustain liquid water on its surface \citep{1993Icar..101..108K}. However, habitability depends on a more complex set of processes, such as the space environment, atmospheric processes, and geological processes \citep[see, e.g.,][]{2016PhR...663....1S,Perryman_2018}. More specifically, we should assess the planet's ability to maintain its atmosphere over its lifetime. The retention of the planetary atmosphere itself depends on a number of processes \citep{2020JGRA..12527639G}. It is commonly assumed that a strong intrinsic magnetic field can act as a shield for the planetary atmosphere, protecting it from the energetic solar particles and solar wind erosion \citep[see reviews by][]{2021SSRv..217...36R,2024RvMG...90..375B}. However, a strong field does not necessarily equate to atmospheric retention, as charged particles can still be stripped away at the open field regions of the planet \citep[i.e., the "polar wind"][]{1968JGR....73.6846B}. Additionally, it has been suggested that a strong planetary magnetic field could drive stronger escape via wave-particle interaction \citep{2010AGUFMSM33B1893S}.

The upper parts of the planetary atmosphere interacts with the space environment created by their host star. A main impact is the ionization of the upper atmosphere by the ionizing radiation --- the star's Extreme Ultraviolet (EUV) and X-ray radiation. Thus, the upper atmosphere refers sometimes as the parts of the atmosphere that are significantly ionized \citep[in contrast to the parts that are considered neutral,][]{Schunk_Nagy_2009}. Many exoplanets, especially habitable M-dwarf exoplanets, are found very close to their host stars, where they are subject to more intense EUV radiation than the conditions around Earth. It has been suggested that this intense radiation could drive a strong hydrodynamic escape \citep[Photoevaporation and the "energy-limited" escape, e.g.,][]{watson_dynamics_1981,Lammer2003,Tian2005extra,Schneiter2007,Penz2008,Murray-Clay2009,Owen2012,Tripathi2015,salz2016energy,Owen19}. Moreover, planets are not guaranteed to remain in the same orbit for the entirety of their existence. For example, the planets in the TRAPPIST-1 system are suspected to have formed at different orbital distances than the ones they are currently observed \citep{Gillon_2017}. If that is the case, then it is possible that some exoplanets that are characterized as habitable may very well only be habitable temporarily until the conditions change over time.

The study of hydrodynamic escape from exoplanets has been focused on hydrogen-based atmospheres of hot jupiters and hot neptunes. These studies were driven by some observational indication for a strong neutral Hydrogen escape, which was supposedly first observed in the Ly$\alpha$ \cite{2003Natur.422..143V}, and the existence of a "Neptunian Desert" \citep[e.g.,][]{Mazeh16,Matsakos16,Lundkvist16,Owen18}. Atmospheric escape from terrestrial planets, which hold a secondary atmosphere made of heavier species, seem to be dominated by non-thermal processes \citep{[see review by][]2020JGRA..12527639G}. Specifically, the escape is carried mostly by heavier ions and not by neutral Hydrogen. In the solar system, $O^+$ escape has been measured for the Earth, Venus, and Mars, and it was found to be in the range of $10^{24}-10^{26}~s^{-1}$ \citep{2021SSRv..217...36R}. Notably, the escape rate of $O^+$ seems to be similar for the three planets, even though they are very different from each other in size, their atmospheric composition, and in their internal magnetic field. Understanding why this is so should play a key role in understanding ion escape from planets and the possible role of the planetary magnetic field in protecting the atmosphere. As a first step, it is important to understand how the increased EUV radiation impacts potentially habitable atmospheres in order to better understand the impact on their escape, and to inform future exoplanet atmosphere surveys. 

In this paper, we use the Global Ionosphere and Thermosphere Model \citep[GITM,][]{RIDLEY2006839} to study the impact of enhanced EUV flux on the upper atmosphere (the top, ionized part) of an Earth-like exoplanet, and what it might mean for habitability of such planets. We focus on the escape of ion species of $O^+$ and $N^+$, as well as the overall structure of the upper atmosphere. Previous work done has largely used one-dimensional models, which are inexpensive to run, and they provide a tool to study many cases. The main drawback of these models in the context of atmospheric escape, is that they do not include horizontal dynamics that may be very important, and could potentially change the escape pattern \citep{2020AGUFMP007.0012B}. GITM allows simulations in 1D or 3D, while using the exact same parameter setting. One goal in this study is to take advantage of this feature to quantify the possible differences between the one- and three-dimensional solutions.  

In Section \ref{sec:model}, we describe our modeling approach and methodology. In Section \ref{sec:results} we present the results for the parameter space examined and compare them to a typical Earth case. We discuss those results in Section \ref{sec:disc} and conclude our findings in Section \ref{sec:conc}.

\section{Model Description} \label{sec:model}

\subsection{General Model Description}
\label{ModelDescription}

The results in this paper were obtained using the GITM model. All the simulations were performed on the MIT SuperCloud cluster \citep{reuther2018interactive}.

GITM is a self-consistent, multi-fluid, 3D model for the upper atmosphere that solves the coupled continuity, momentum, and energy equations for neutrals and major ion species, and the interaction between them. The equations are solved on a spherical grid, and the coordinate system is height-based rather than pressure-based. The equations include a number of source terms that capture additional physics, which is not included in the ideal set of equations. These source terms are: 1) chemistry in the continuity equation (photoionization, creation and destruction of species); 2) ion drag, viscosity, and gravity wave acceleration for the horizontal momentum; 3) ion drag and friction between neutrals for vertical momentum; 4) radiative cooling, EUV heating, auroral heating, Joule heating, heat conduction, and chemical heating for the neutral temperature equation. The output of the model includes several neutral and ion species densities, their associated velocities, temperatures, electron densities, conductance, and heating/cooling rates of the processes included in the model. GITM has been vastly use to study the upper atmosphere of the Earth \citep[e.g.,][]{}, Venus \citep{2008JGRA..113.9302D,2011JASTP..73.1840P,2011JGRA..11612305Y,2015JGRA..120.1248S,2024JGRE..12908079P}, Mars \citep[][]{2015JGRE..120..311B,2015JGRA..120.7857D,2023JGRE..12807670P}, and Titan \citep{2010JGRE..11512002B,2011JGRE..11611002B}. We refer the reader to \cite{RIDLEY2006839} and the references above for the complete model description. 

\subsection{EUV Input}
\label{EUVflux}

By default, the model sets the input EUV flux is via the $F_{10.7}$ flux. This value is a measure of the solar $10.7~cm$ flux and is given in solar flux units, with 1 sfu = $10^{-22}~Wm^{-2}~Hz^{-1}$. The $F_{10.7}$ flux is generally used as a proxy for the solar EUV output, and GITM uses the SERF model\citep{1981GeoRL...8.1147H,1991JATP...53.1005T} and EUVAC to convert the $F_{10.7}$ flux into an EUV spectrum of 59 wavelength bins. Given the $F_{10.7}$ flux, the SERF model determines the intensity of each wavelength by assigning a corresponding weight value for the respective wavelength. As an alternative, GITM also allows to provide a detailed input EUV spectrum with a similar binning as an input file to the model \citep{2011JASTP..73.1840P}. However, in this initial study we choose to use the $F_{10.7}$ flux as an input so that we can focus on simply multiplying our reference spectrum for a baseline case by a constant. Thus, the usage of a more detailed input EUV spectrum, including that of an M-dwarf star, is left for a follow up investigation. Once the EUV input is defined, ionization and heating rates are then determined by the given cross sections in \cite{1979JGR....84.3360T}. 

\subsection{Grid Structure}
\label{Grid}

The vertical grid is assigned based on the planetary scale height, which can be calculated feeding the provided $F_{10.7}$ parameter into the Mass Spectrometer Incoherent Scatter radar (MSIS) model \citep{MSIShttps://doi.org/10.1029/2020EA001321}. This model describes atmospheric temperature and number densities for select species based on input parameters. That temperature and the number densities can then be used to determine the scale height, given by:

\begin{equation}
    h = \frac{k_BT}{mg},
\end{equation}

\noindent where $k_B$ is the Boltzmann constant, $T$ is temperature, $m$ is mean molecular mass at that height, and $g$ is local gravity. However, MSIS is a empirical model that is designed for typical Earth conditions and quickly breaks down for the more extreme values of $F_{10.7}$ which are used in this study. Thus, this empirical model needs to be bypassed. Here, we define the vertical grid by assigning an analytical function for the initial temperature profile. This profile then sets the local scale-height and the overall vertical grid structure. The model defines input parameters $T_{max}$, $T_{min}$, $r_{TW}$, and $r_{TH}$. These parameters specify, respectively, the initial temperature at the top and the bottom of the thermosphere, the vertical width of the temperature gradient, and the midpoint height of the temperature profile. These parameters are used to define the initial temperature profile used by the model to create the vertical grid: 

\begin{equation}
    T_{avg} = \frac{T_{max}+T_{min}}{2}
\end{equation}
\begin{equation}
    T_{diff} = \frac{T_{max}-T_{min}}{2}
\end{equation}
\begin{equation}
    T = T_{avg} + T_{diff}\cdot \mathrm{tanh}\left(\frac{r-r_{TH}}{r_{TW}}\right)
\end{equation}

We manually define the lower boundary altitude, and the scale height for that altitude is calculated. The vertical size of that grid cell is set at 0.3 scale heights for most runs. However we decrease this factor to 0.2 for the extreme EUV cases. The model then calculates the scale height at the top of that grid cell and uses it to determine the size of the next grid cell, repeated for each grid cell until the total number of vertical grid points is achieved (defined by the user), so the entire vertical grid has been set. Each model run described here uses a lower boundary altitude of 100 km. The maximum height is determined by the combination of the scale heights, and the number of radial (altitude) grid points. 

The model has a number of optional drivers and source terms which may be included depending on the goals of those who utilize it. We use a constant solar wind with a velocity of $400~km~s^{-1}$ and an interplanetary magnetic field of $B_z=-2.0~nT$. We do not use auroral drivers. Neutral heating efficiency is set to 0.05 and photoelectron efficiency is 0.0. We use an Eddy diffusion coefficient of 50.0. 

In this study, we include both 1D and 3D solutions for various values of the $F_{10.7}$ flux input. For 3D cases, the horizontal grid spacing is defined manually by specifying the bounds of latitude and longitude along with the number of grid cells to use in each direction. The model allows for the user to use a stretched latitudinal grid spacing if desired, but we use a uniform latitudinal grid in this study. For the 1D cases, a constant position of longitude=180, latitude=0 is used for subsolar cases, and longitude=180, latitude=80 is used to capture polar data. The longitudinal grid is always uniform and is the same for all runs. For the 3D cases, we use a full horizontal grid with dimensions 80 by 48 grid cells, spanning the full 360 degrees in longitude and 180 degrees in latitude, respectively, resulting in a resolution of 4.5 and 3.75 degrees in longitude and latitude, respectively. 

\subsection{Scaling the EUV Flux}
\label{ScalingEUV}

We use a baseline, fiducial, averaged solar value of 150 for the $F_{10.7}$ flux. In this study we treat this value as the baseline case, or an EUV factor of 1. For the 1D cases, we use EUV factors of 1, 5, and 9. For the 3D cases, we use enhanced fluxes with EUV factors of 1, 3, 5, 7,  9, 10, 20, and 50 times the baseline value. For each case, the corresponding $F_{10.7}$ flux value is used as input into the model. The only other changes made for each run are the grid cells via the $T_{max}$ input parameter, which is increased to allow for increased vertical reach of the model to capture relevant information that is present higher in the atmosphere for the higher EUV scale factors. While the final result has some minimal dependency on the value of $T_{max}$ (mostly convergence time), we find that the EUV scale factor dominates the results as far as comparing them with the baseline 1x scale factor case. By default, the model uses a numerical factor of 0.3 on the scale height when defining the vertical grid size. For EUV factors 10 and less, we use the default 0.3 value. However, for EUV factors of 20 and 50, we reduce this value to 0.2 to increase the vertical resolution, a necessary change to avoid the model producing unphysical results. The scale height factors, combined with the number of vertical grid cells are set to achieve similar maximum altitude for the different EVU flux cases (the overall height of the atmosphere is different). Simulations for EUV factors of less than 20 are allowed to reach steady state solutions. EUV factors of 20 and 50 do not reach steady states for reasons that will be presented in our discussion. Results for scale factors 1--10 are produced by running the model from March 12, 2002 to March 21, 2002. Results for scale factors 20 and 50 are produced running the model from March 1, 2002, to March 21, 2002. 


\section{Results} \label{sec:results}

\begin{figure*}
    \centering
    \includegraphics[width=0.5\linewidth]{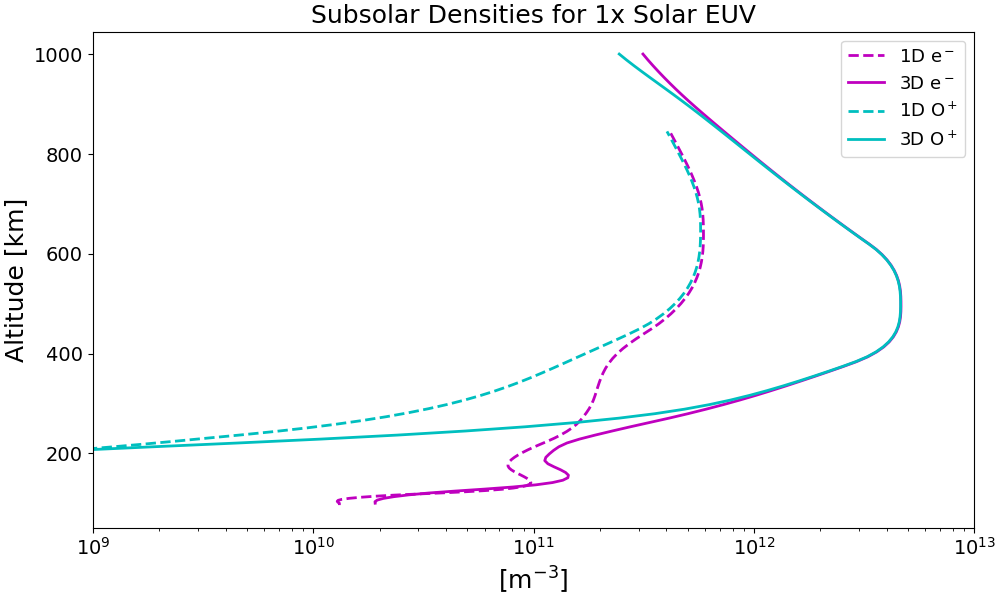}\hfill
    \includegraphics[width=0.5\linewidth]{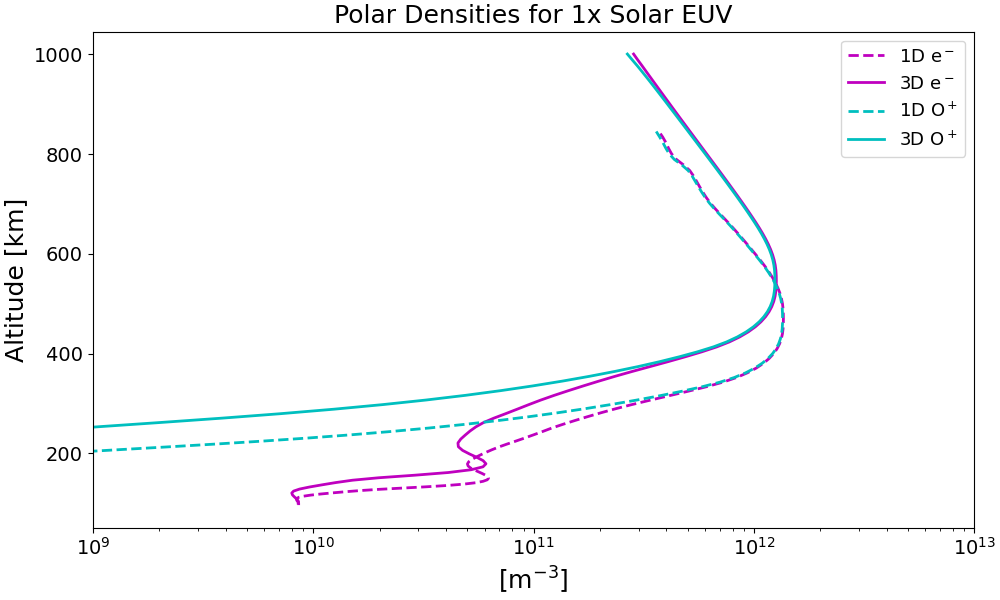}
    \includegraphics[width=0.5\linewidth]{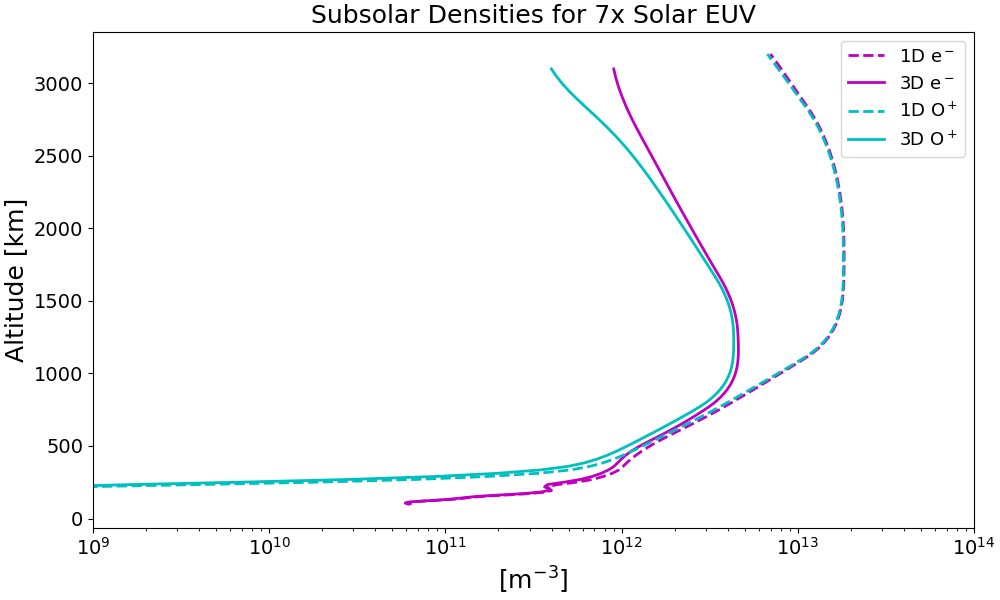}\hfill
    \includegraphics[width=0.5\linewidth]{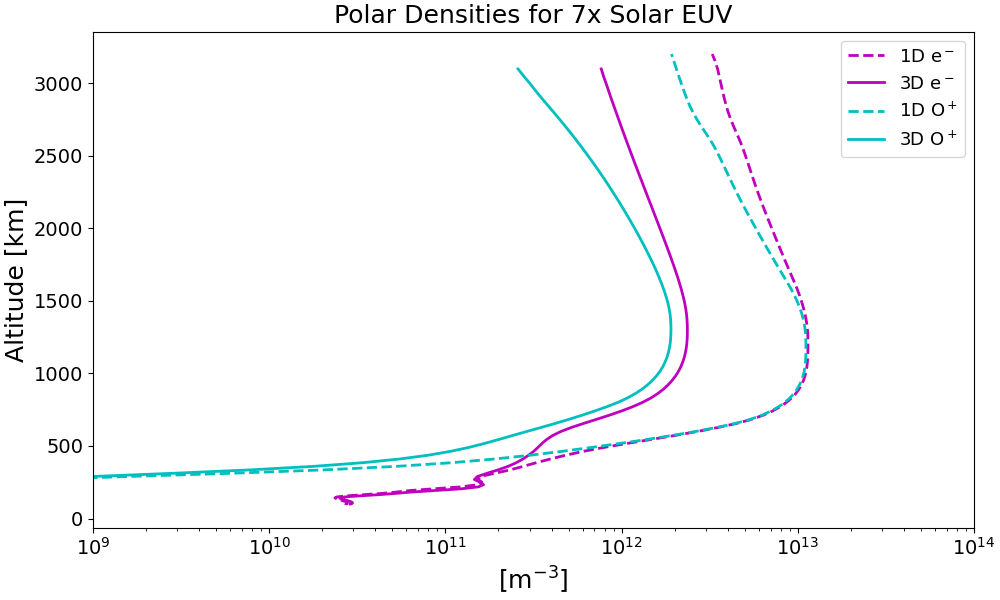}
    \includegraphics[width=0.5\linewidth]{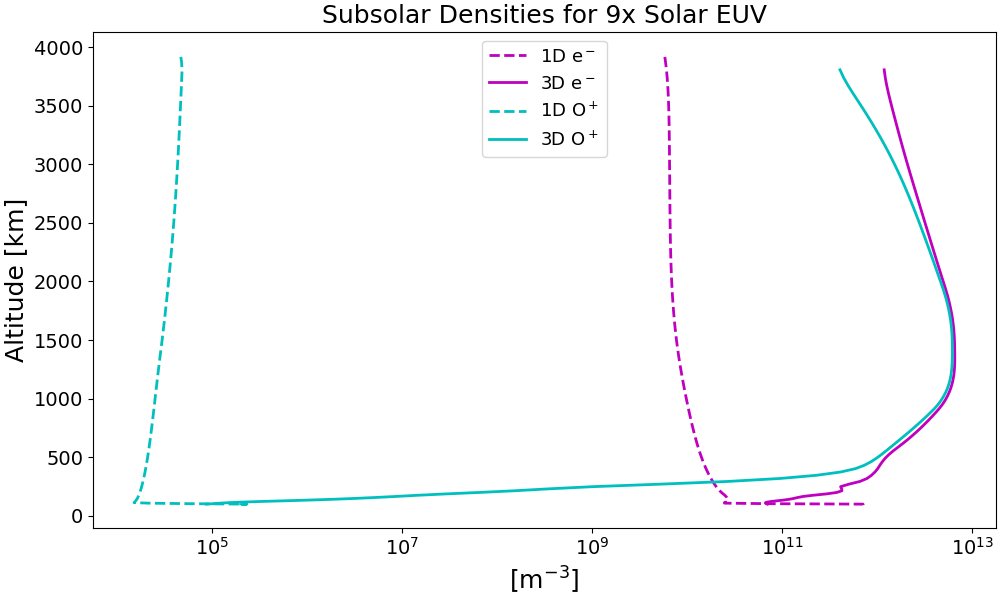}\hfill
    \includegraphics[width=0.5\linewidth]{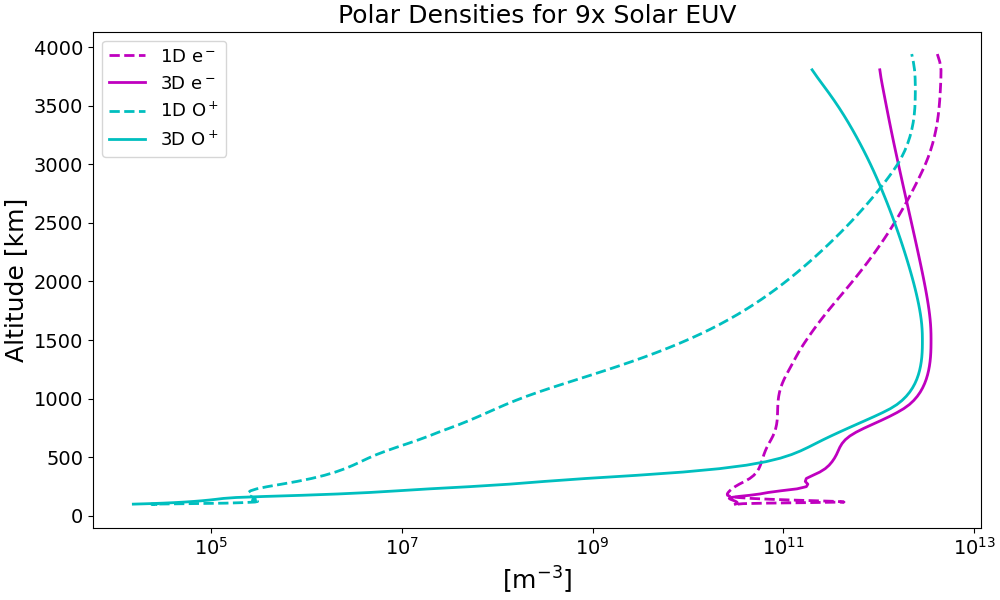}
    \caption{1D and 3D vertical density profiles for electrons and $O^+$ for different solar EUV input.}
    \label{fig:1d3d_density}
\end{figure*}

\begin{figure*}
    \centering
    \includegraphics[width=0.5\linewidth]{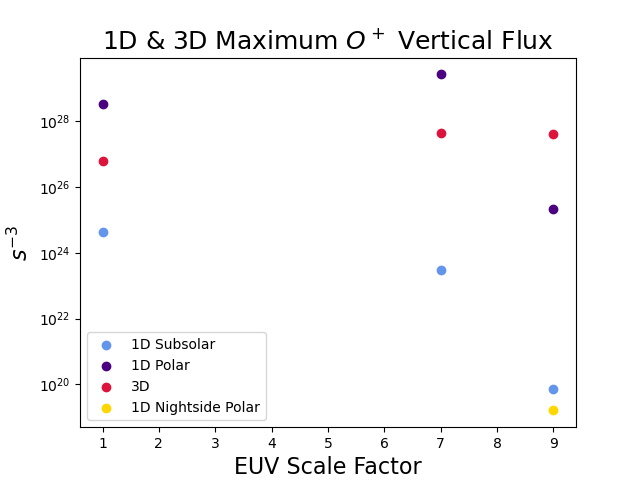}\hfill
    \includegraphics[width=0.5\linewidth]{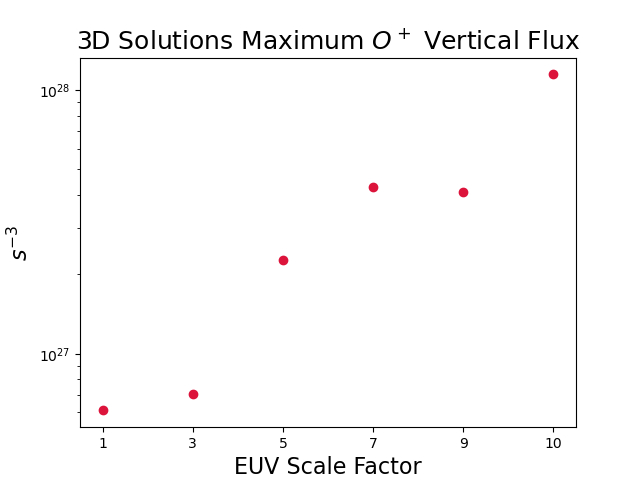}
    \caption{\textit{Left:} Maximum vertical upward flux of $O^+$ for 1D subsolar and polar solutions as well as 3D solutions for scale factors 1, 7, and 9. \textit{Right:} Maximum vertical upward flux of $O^+$ for all 3D solutions with a scale factor less than or equal to 10. }
    \label{fig:1d3d_escape}
\end{figure*}

\subsection{General Density Structure and Comparison Between 1D and 3D Models}
\label{1D3D}

Fig \ref{fig:1d3d_density} shows the density structure for electrons and $O^+$. The results are for the 1x, 7x, and 9x solar EUV levels, using both 1D and 3D modes. The 1D model is calculated at two selected locations --- one at the subsolar point and the other at the day side near the North Pole (180 degrees longitude, 80 degrees north of the equator). The first notable feature is that the peak of the density profile (both electrons and $O^+$) increases by an order of magnitude with increasing EUV flux. The 3D solution shown here is extracted at the same locations as the 1D cases. The 3D, subsolar solution for the 1X case agrees quite well with the profile found in the literature for the Earth \citep[e.g,][]{Schunk_Nagy_2009}. Another notable feature is the difference between the 1D and 3D solutions. For the lower EUV factors of 1 and 7, the polar solutions are closer, but they differ by an order of magnitude at the equator. For the intense, EUV factor of 9 case, the solutions differ even more. Moreover, the 3D solutions in all cases seem more physical in shape compared to the 1D solutions. Further analysis of the 1D solutions for this EUV factor of 9 shows a lack of convergence in the solution, likely stemming from the forcing of all particles into a vertical column and leading to the 1D simulations for this case to behave more closely to the extreme EUV cases. The results presented in Fig \ref{fig:1d3d_density} show that 1D models could be limited in capturing the effects that are covered by 3D models, and they may produce less physical results, especially for more intense EUV radiation which is anticipated for close-in exoplanet stellar environments. Moreover, 1D models could produce very different results depending on the selected location they represent. Thus, all of these consequences should be accounted for when 1D models are used to simulate planetary atmospheres. 

Fig \ref{fig:1d3d_escape} shows the maximum O$^+$ upward vertical mass flux, $\Phi$ in [$s^{-1}$], for the two 1D solutions and the 3D solution for EUV  factors of 1, 7, and 9. We also performed an additional 1D solution for the 9x case at 80N, 0W. For the 1D solutions, the flux is calculated assuming a spherical symmetry on a sphere with radius $r$:
\begin{equation}
    \Phi \left( r \right) = 4\pi r^2 \rho(r) v(r)
    \label{eq:1dverticalflux}
\end{equation}
where $\rho$ is the species's mass density, $v$ is the species's vertical velocity, and $r$ is the distance from the geometric center of the planet (Earth, in this case). For the 3D solutions, the flux is calculated by integrating (summing) the local mass flux over the horizontal grid at radius $r$:
\begin{equation}
    \Phi \left( r \right) = r^2 d\phi d\theta \sum_{i,j} \rho_{ij}(r) v_{ij}(r) \sin \theta_{i}.
    \label{eq:3dverticalflux}
\end{equation}
Here, the sum over $i$ and $j$ integrates over all the horizontal grid cells (latitude and longitude) at a given altitude, $\theta_i$ is the latitude, and $d\phi$ and $d\phi$ represent the angular resolution of each horizontal grid cell (uniform for this study). 

Just like the results in Fig \ref{fig:1d3d_density}, the left panel of Fig \ref{fig:1d3d_escape} shows notable difference between the 1D and 3D solutions. For lower EUV factors, the polar 1D solution overestimates the 3D one, and the subsolar solution underestimates it. Differences are 1-2 orders of magnitude. For an EUV factor of 9, all three 1D models underestimate the flux compared to the 3D solution. This again highlights that 1D solutions are limited in providing a global picture. In the right panel of Fig \ref{fig:1d3d_escape} we present the maximum vertical $O^+$ flux for each of the EUV scale factors that are less than or equal to 10 (only 3D solutions). We observe that the baseline solution is on the order of $10^{26}~s^{-1}$, which is the same order of magnitude of the observed flux at Earth and it is not off by many orders of magnitude \citep[e.g.,][]{2020JGRA..12527639G,2021SSRv..217...36R}. While this is not necessarily a quantitative validation of the model for the Earth due to the fact that GITM does not include the polar wind process, it provides some reasonable assessment for the magnitude of the ion escape for the baseline case. We also note that GITM does not have the ability to track particles as they leave the upper boundary of the grid so we cannot distinguish between particles who will eventually return or leave the planet. These limitations are currently unavoidable, and so we use the tools at our disposal to examine the overall trends compared to the baseline case, which is accurate enough to capture large scale features of the upper atmosphere. As the scale factor increases, we see that the vertical flux also increases in magnitude, with the EUV factor of 10 showing a vertical flux roughly two orders of magnitude greater than the baseline case.


\begin{figure*}[ht!]
    \centering
    \includegraphics[width=0.5\linewidth]
    {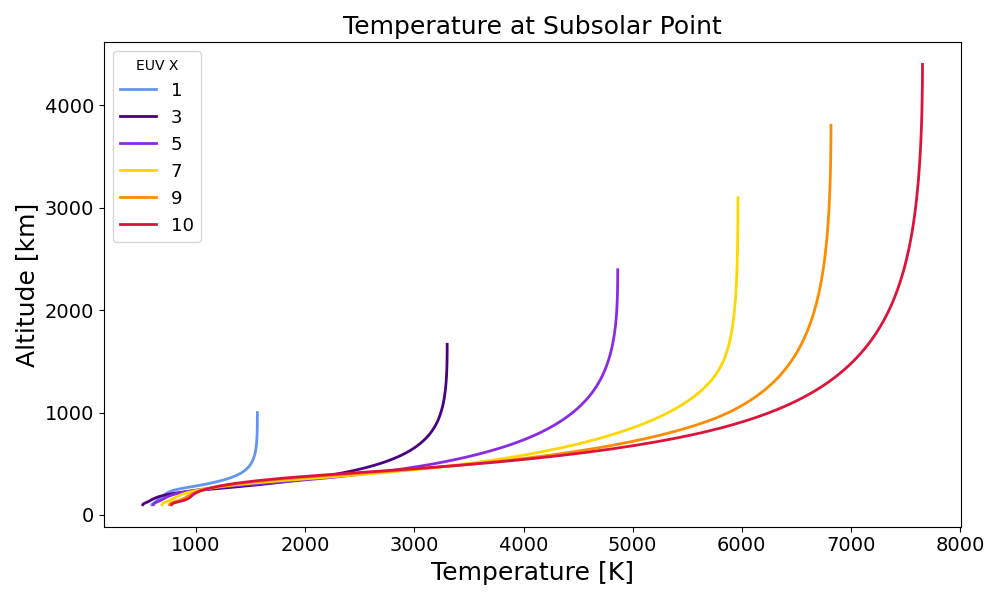}\hfill
    \includegraphics[width=0.5\linewidth]{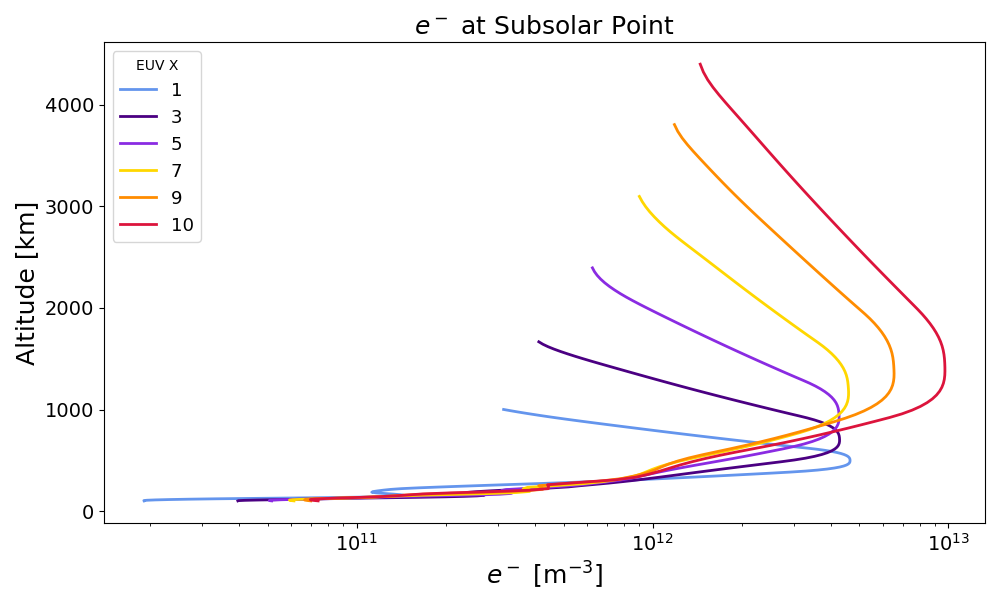}
    \includegraphics[width=0.5\linewidth]{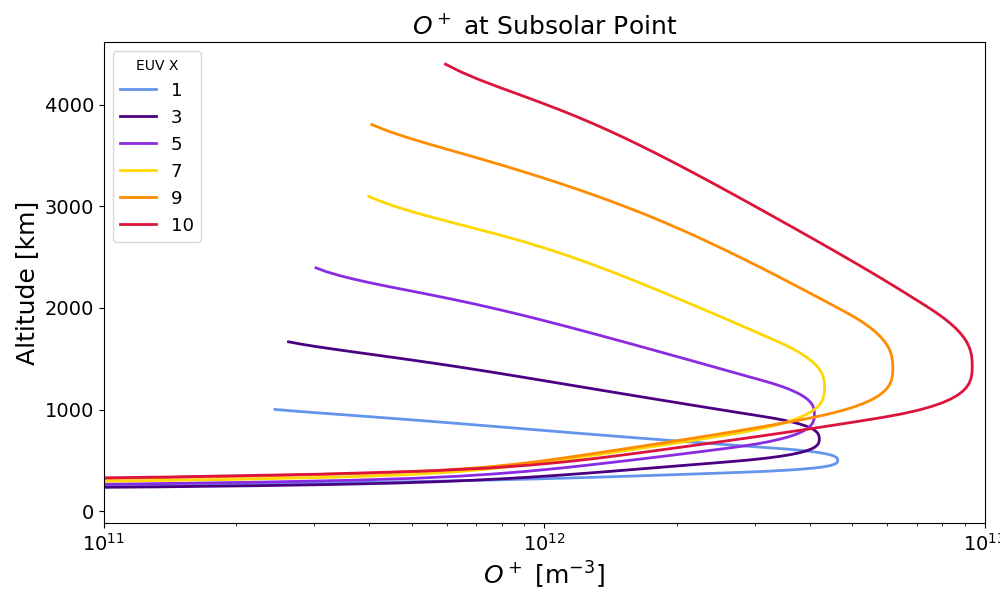}\hfill
    \includegraphics[width=0.5\linewidth]{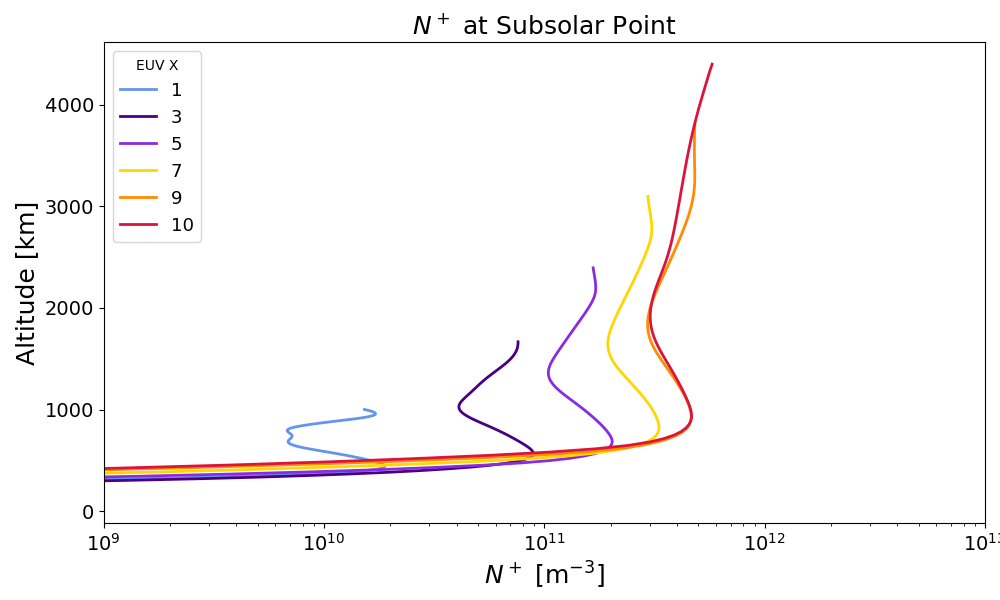}
    \caption{Subsolar point comparison for all EUV irradiance levels, showing temperature (top left), electron density (top right), O$^+$ density (bottom left), and N$^+$ density (bottom right).}
    \label{fig:daysidecomps}
\end{figure*}


\begin{figure*}[ht!]
    \centering
    \includegraphics[width=0.3\textwidth]{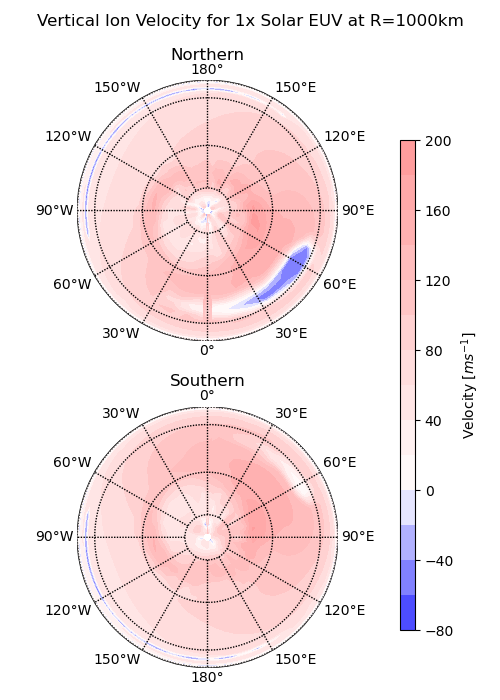}\hfill
    \includegraphics[width=0.3\textwidth]{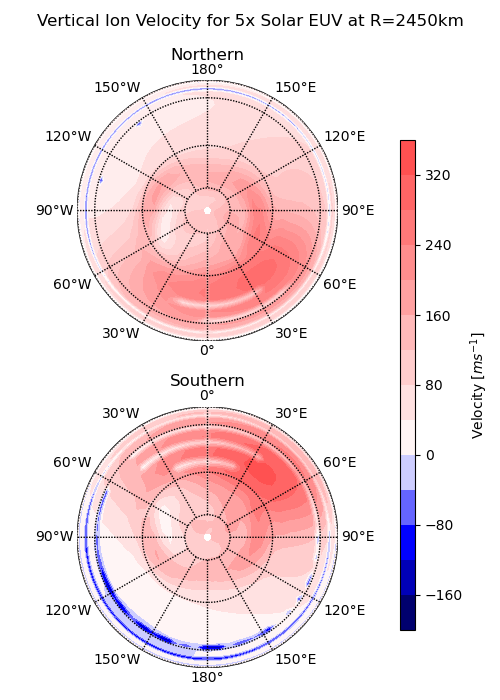}\hfill
    \includegraphics[width=0.3\textwidth]{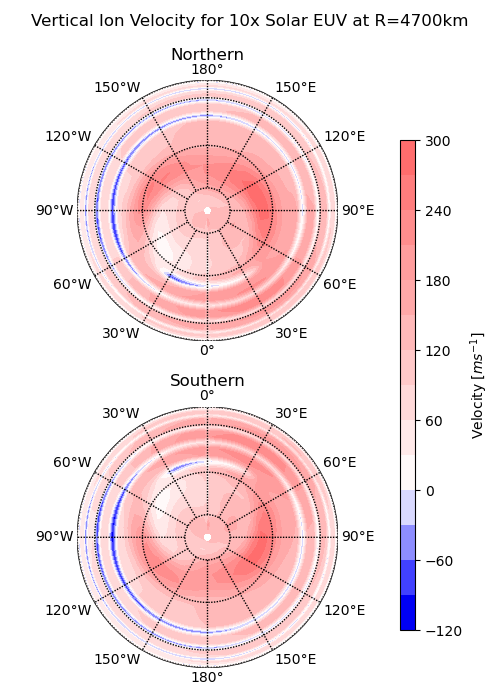}\hfill
    \caption{Vertical ion velocities for EUV scale factors 1, 5, and 10. Altitudes are chosen to align roughly with the maximum vertical flux of $O^+$ shown in Fig \ref{fig:1d3d_escape}.}
    \label{fig:vionup}
\end{figure*}

\begin{figure*}[ht!]
    \centering
    \includegraphics[width=0.3\textwidth]{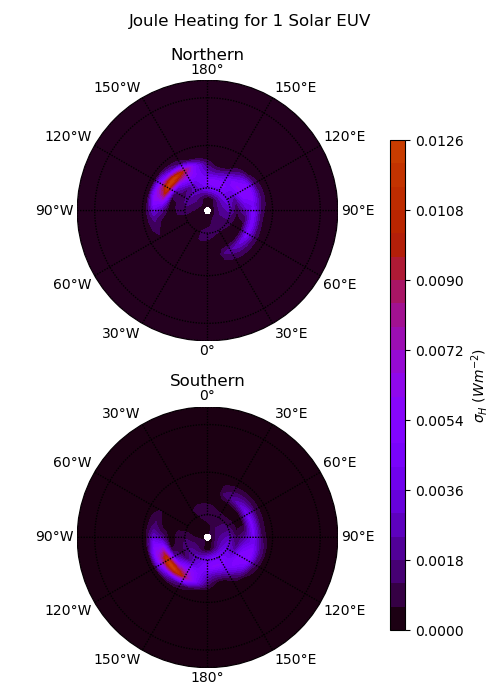}\hfill
    \includegraphics[width=0.3\textwidth]{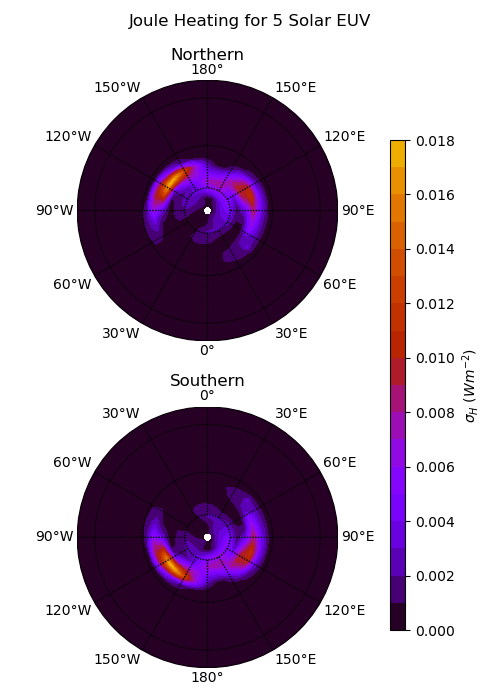}\hfill
    \includegraphics[width=0.3\textwidth]{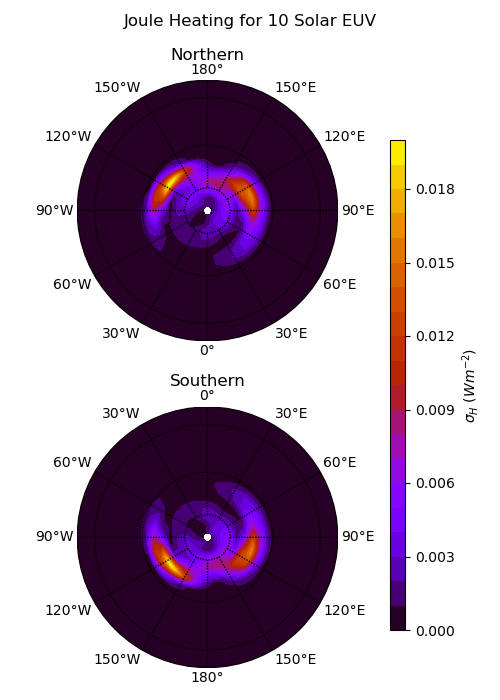}\hfill
    \caption{Altitude integrated joule heating for scale factors 1, 5, and 10. Gradient uses the same maximum value for visible demonstration of the changes between the images}
    \label{fig:jouleheating2D}
\end{figure*}

Fig \ref{fig:daysidecomps} shows the vertical density profiles for electrons, $O^+$, and $N^+$, as well as the atmospheric temperature profile at the subsolar point of each 3D solution. For EUV scale factors 1, 3, 5, and 7, we observe relatively little change in the peak subsolar magnitude of electrons and $O^+$, however for scale factors 9 and 10 we observe that the height at which the peak occurs changes very little while the density increases significantly. Additionally, we observe that the $N^+$ density increases for increasing EUV scale factor, until the scale factor reaches 9-10, which show little variation in their vertical density profiles. This would suggest that the density of $N^+$ becomes supply limited rather than ionizing energy limited. These density profiles suggest a transition of sorts at the higher EUV levels simulated with the scale factors of 9 and 10. 

\subsection{Ionic Oxygen Vertical Flux and Heating}
\label{O+_heating}

In order to better understand our results, we examine the global vertical ion velocity for the different EUV factors. This is shown in Fig \ref{fig:vionup}. For each case, we plot the horizontal distribution at the selected altitude that corresponds to the maximum vertical $O^+$ flux values presented in Fig \ref{fig:1d3d_escape}. One can see that the strongest upward vertical motion occurs in the polar regions, where most heavy ion escape follows open magnetic field lines. It can also be seen that the maximum upward vertical velocity increases with the increase of the EUV factor.

Fig \ref{fig:jouleheating2D} shows the altitude-integrated Joule heating data for different EUV factors up to 10. The majority of the heating occurs around the poles, and we see an increase in the magnitude as the EUV intensity increases. 

\begin{figure*}[t!]
    \centering
    \includegraphics[width=0.5\linewidth]{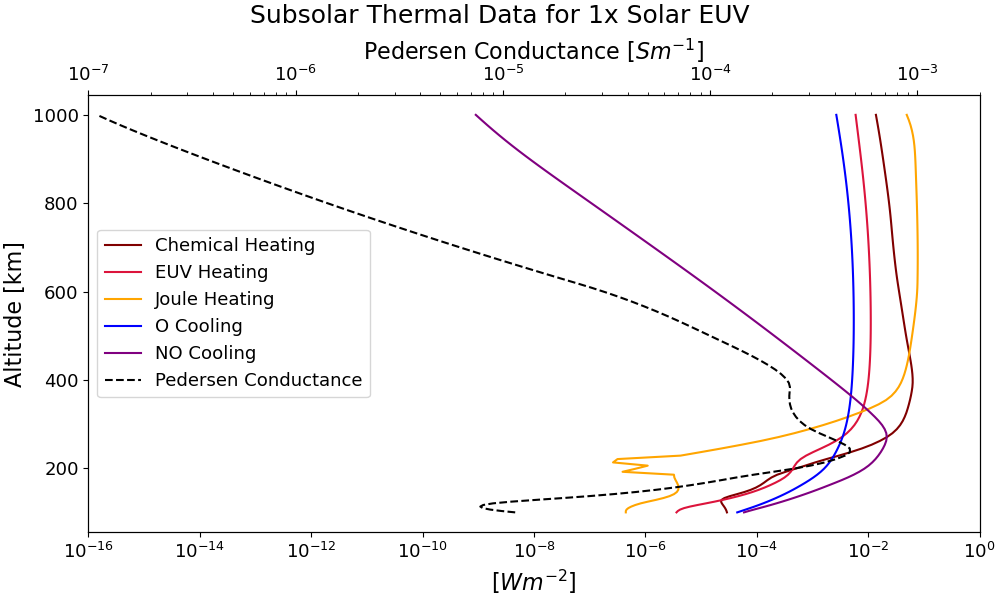}\hfill
    \includegraphics[width=0.5\linewidth]{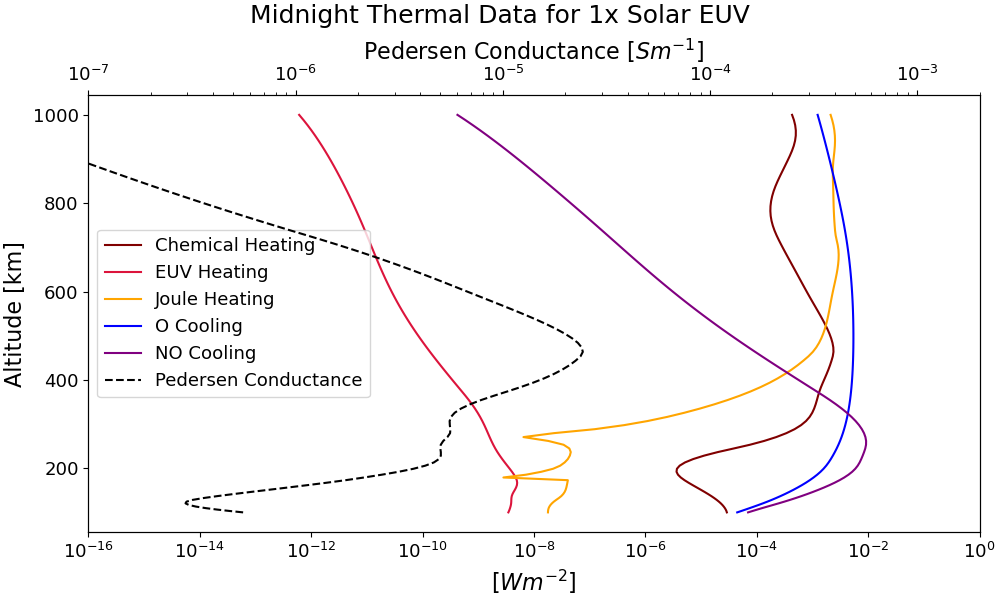}
    \includegraphics[width=0.5\linewidth]{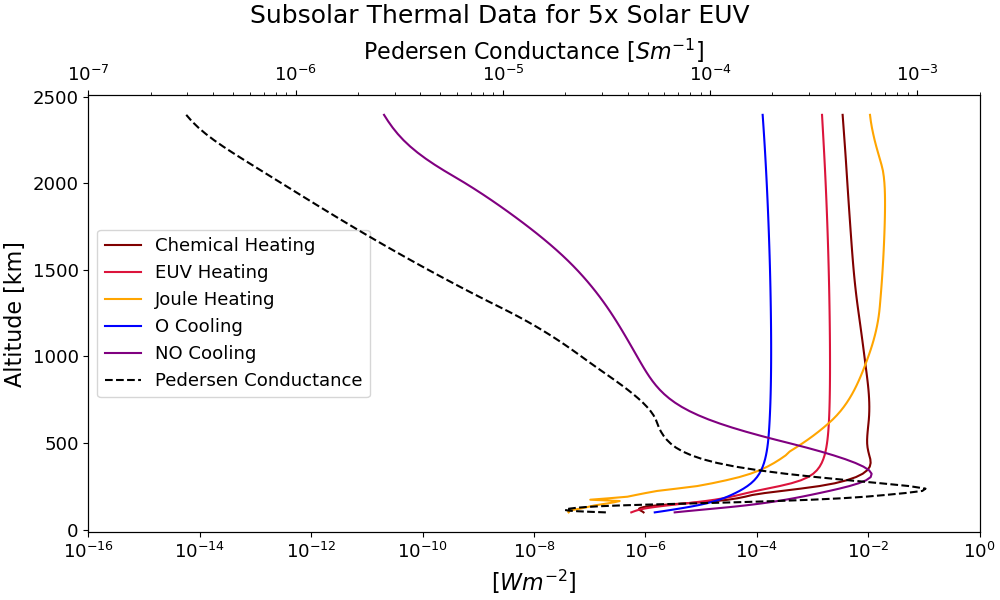}\hfill
    \includegraphics[width=0.5\linewidth]{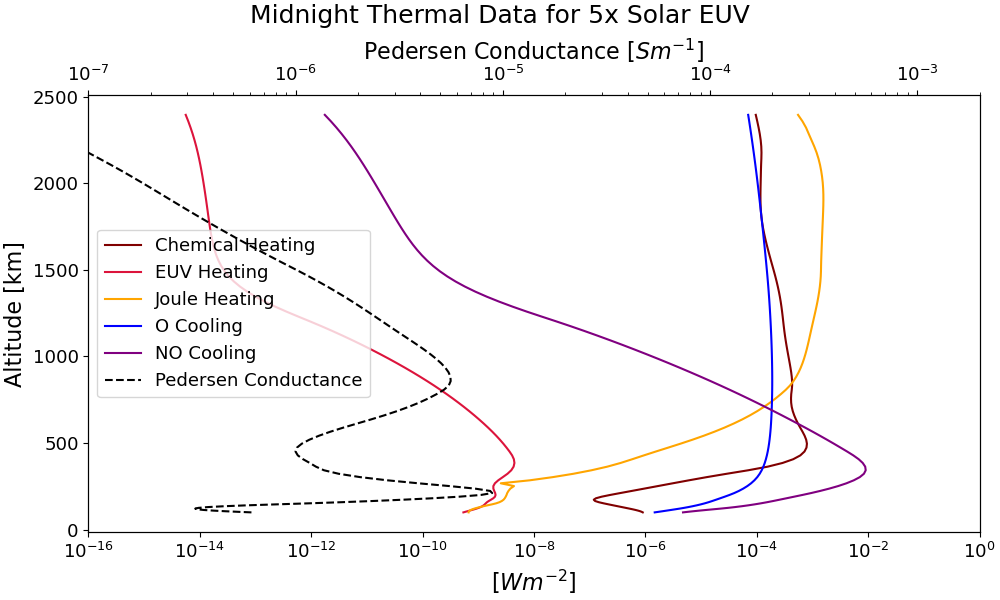}
    \includegraphics[width=0.5\linewidth]{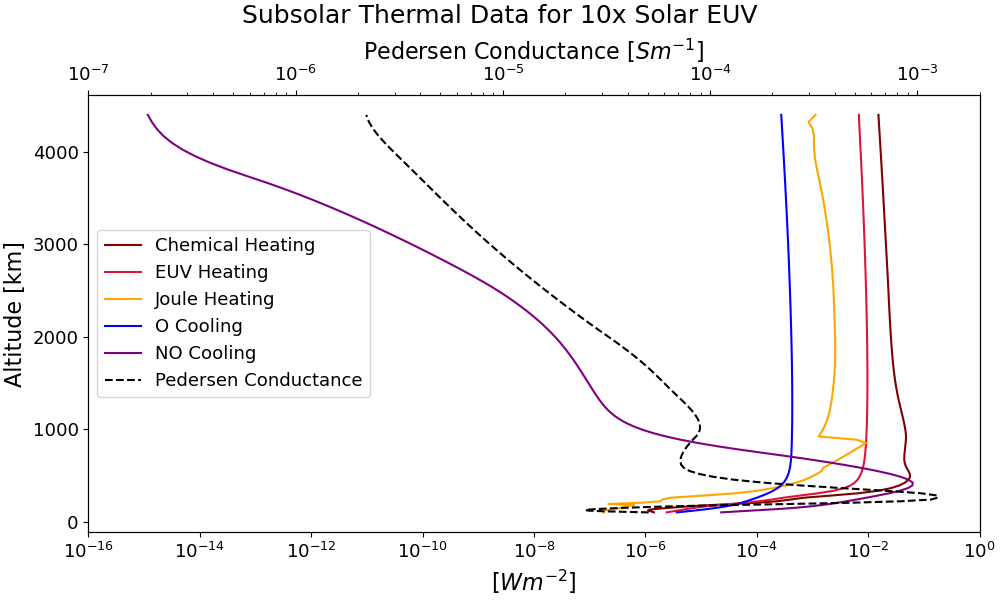}\hfill
    \includegraphics[width=0.5\linewidth]{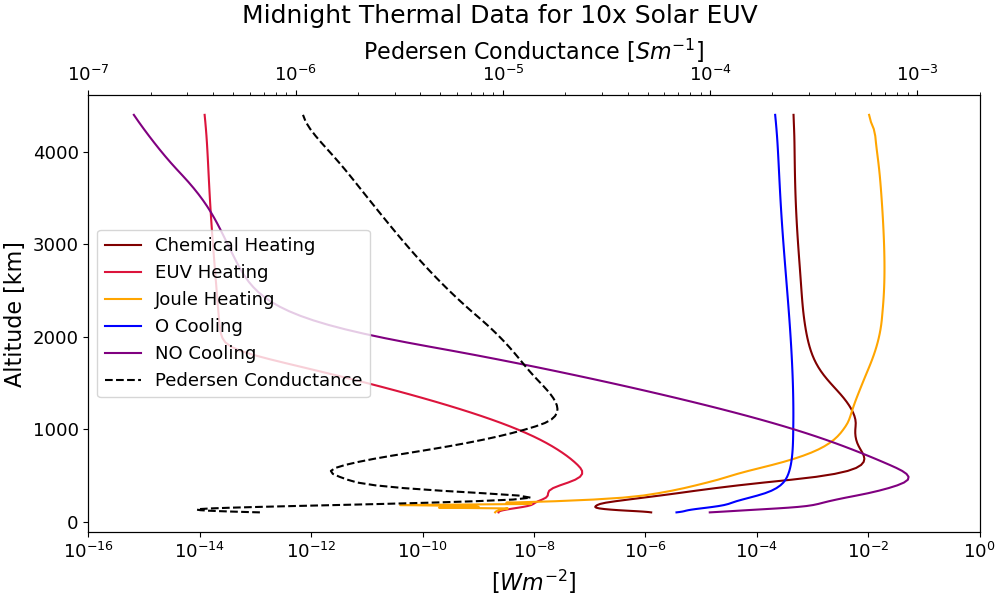}
    \caption{Subsolar and midnight heating and cooling rates along with Pedersen conductance. Cooling rates are multiplied by -1 in order to plot them on the same axes as the heating rates.}
    \label{fig:heatingcooling}
\end{figure*}

\begin{figure*}[t!]
    \centering
    \includegraphics[width=0.5\linewidth]{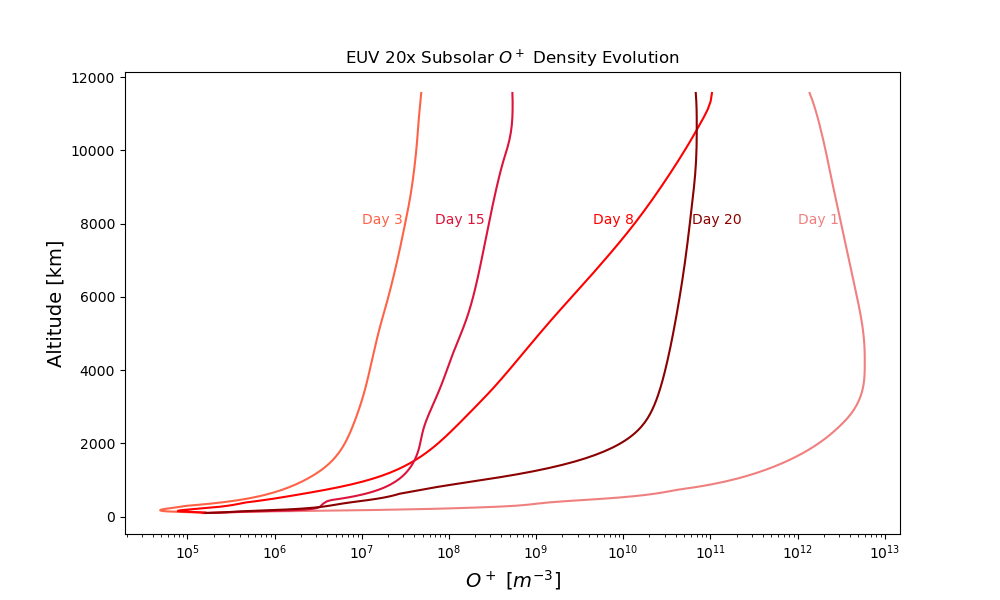}\hfill
    \includegraphics[width=0.5\linewidth]{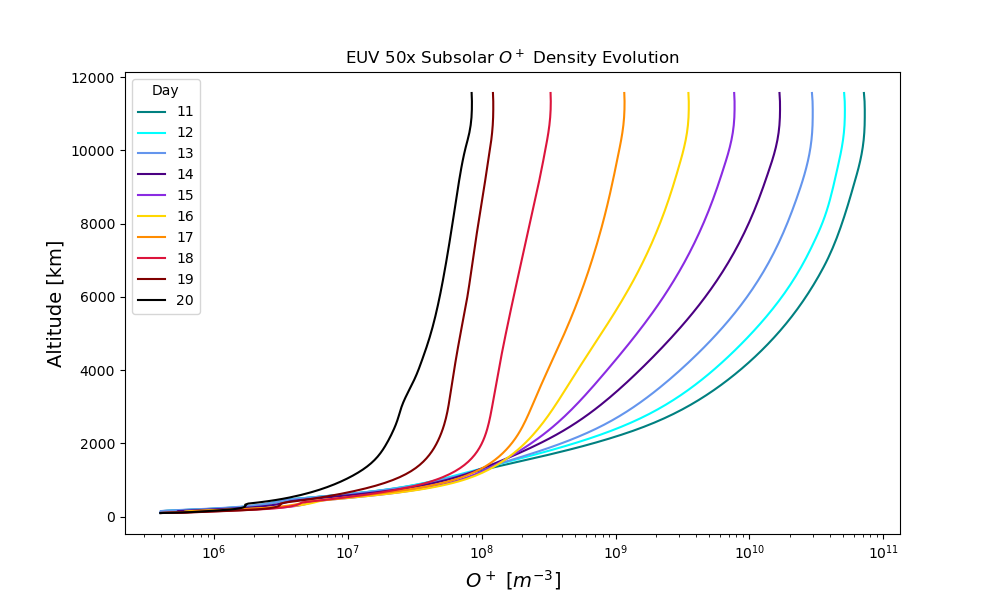}
    \caption{Evolution of subsolar $O^+$ density for extreme EUV inputs.}
    \label{fig:extreme_O+}
\end{figure*}

Fig \ref{fig:heatingcooling} shows heating and cooling rates, as well conductivity as a function of altitude for EUV factors of 1, 5, and 10. Data is extracted from the 3D solutions at the subsolar and midnight points. For all three cases, we see the EUV heating remains relatively constant with altitude once above ~300km. Subsolar Joule heating actually decreases slightly with increasing EUV radiation, a potential result of the increased Pedersen conductance in the higher EUV radiation cases. $O$ cooling for the baseline EUV flux is on the order of 0.01 $Wm^{-2}$, however for the 5x case it decreases to $10^{-4} Wm^{-2}$ and in the EUV factor of 10 case, it is $10^{-3} WM^{-2}$. Dayside $NO$ cooling rate appears to be similar for all cases. Above 1000 km there is a bump where the rate at which the cooling drops off decreases before once again dropping off 4 or 5 orders of magnitude every 1000 km. Pedersen conductance peaks around an altitude of 200 km for all EUV levels, however for the levels above the baseline 1x solar EUV the peak conductance is about double that of the conductance for the baseline case. Additionally, we observe a significant change in the altitude profile of the Pedersen conductance as the EUV flux is increased. In the baseline case Pedersen conductance has decreased to an order of $10^{-7} Sm^{-1}$ at 1000 km. In the EUV factor of 5 case, it is at $10^{-5} Sm^{-1}$ at 1000 km and doesn't drop to $10^{-7} Sm^{-1}$ until 2500 km. In the EUV factor of 10 case, the conductance at 1000 km is even greater than $10^{-5} Sm^{-1}$ and doesn't drop off to $10^{-6} Sm^{-1}$ until altitudes of over 4000 km. On the midnight side, we see an expected drop in the EUV heating with altitude. However, joule heating increases by about an order of magnitude from the baseline case to the EUV factor of 10 case, as does the chemical heating. Nightside $NO$ cooling increases between 1 and 2 orders of magnitude in the lower atmosphere from the baseline case to EUV factor of 10 case but drops off significantly with altitude, even when accounting for the change in vertical scale. At 2500 km the $NO$ cooling of the 5x solar EUV case is on the order of $10^{-12} Wm^{-2}$ and at the same height in the EUV factor of 10 case it has dropped nearly another two orders of magnitude. We see the relative peak of the Pedersen conductance push higher into the atmosphere same as with the day side and occurring at approximately the same height.

\subsection{Extreme EUV Solutions}
\label{extreme}

Our most two extreme cases are EUV factors of 20 and 50 times the baseline EUV levels. In order to capture the altitude profiles of these extreme EUV levels, we increase the size of the altitude grid to span just under two planetary radii (Earth radii). This requires longer convergence times in order to capture the details of the solutions. The resulting solutions differ greatly from the other EUV levels in a few ways. In Fig \ref{fig:extreme_O+} we present the time evolution of the $O^+$ density for the extreme solutions. For the EUV factor of 20 case, we observe a cyclic behavior between atmospheric growth and atmospheric depletion (five selected times are shown). For this level of EUV flux, the atmosphere has enough time to build up, but then it is depleted due to the extreme heating. The cyclic behavior is due to the fact that the atmosphere is continuously fed by ionization. There is enough energy in the upper atmosphere to result in the escape of $O^+$. However, the neutral source at the bottom of the atmosphere provides a constant supply of $O$ and $O_2$ to the upper atmosphere. For the EUV factor of 50 case, we present the final ten days of the simulation, which very clearly depict a continuous depletion of the atmosphere. The supply of neutral $O$ and $O_2$ is not sufficient to replenish the upper atmosphere given the rate at which it is escaping under such extreme conditions.

\section{Discussion} \label{sec:disc}

\begin{figure*}[ht!]
    \centering
    \includegraphics[width=0.5\linewidth]{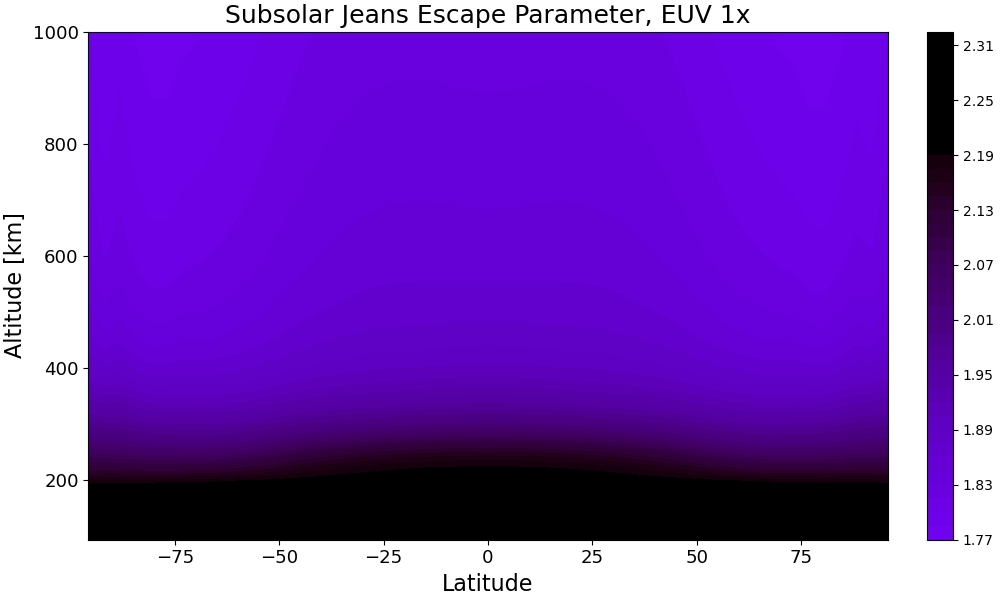}\hfill
    \includegraphics[width=0.5\linewidth]{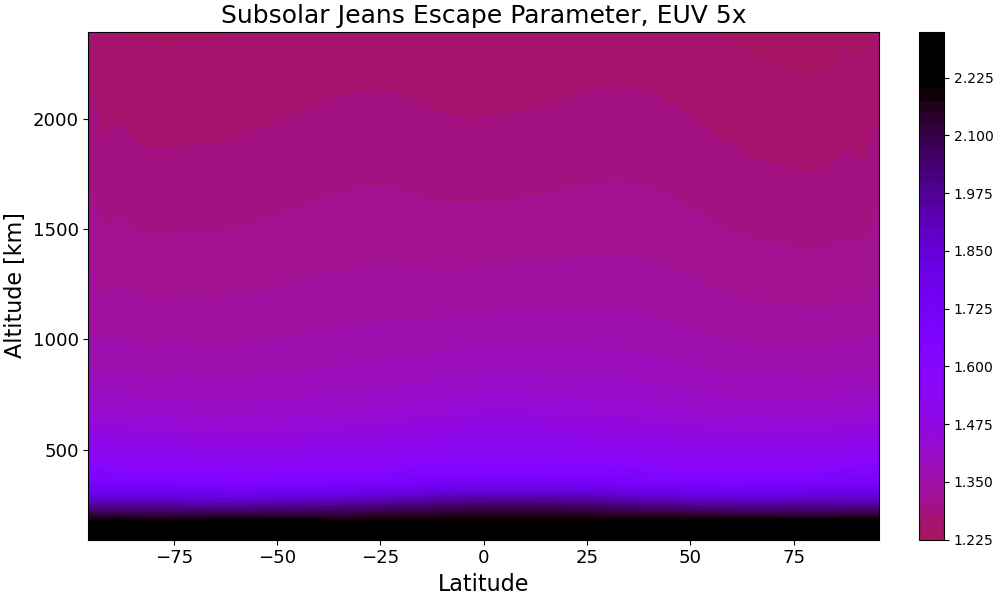}\hfill
    \includegraphics[width=0.5\linewidth]{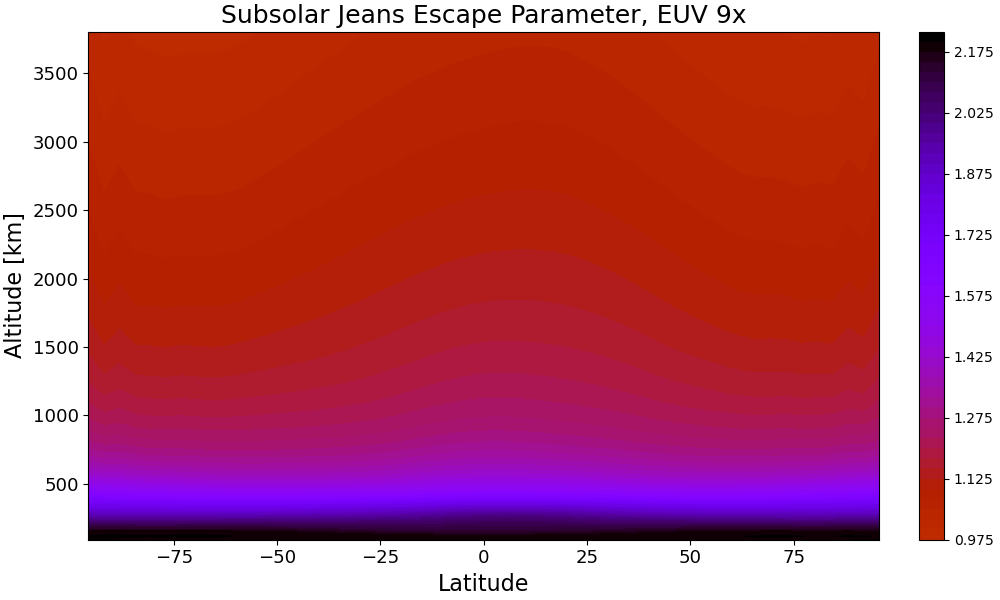}\hfill
    \includegraphics[width=0.5\linewidth]{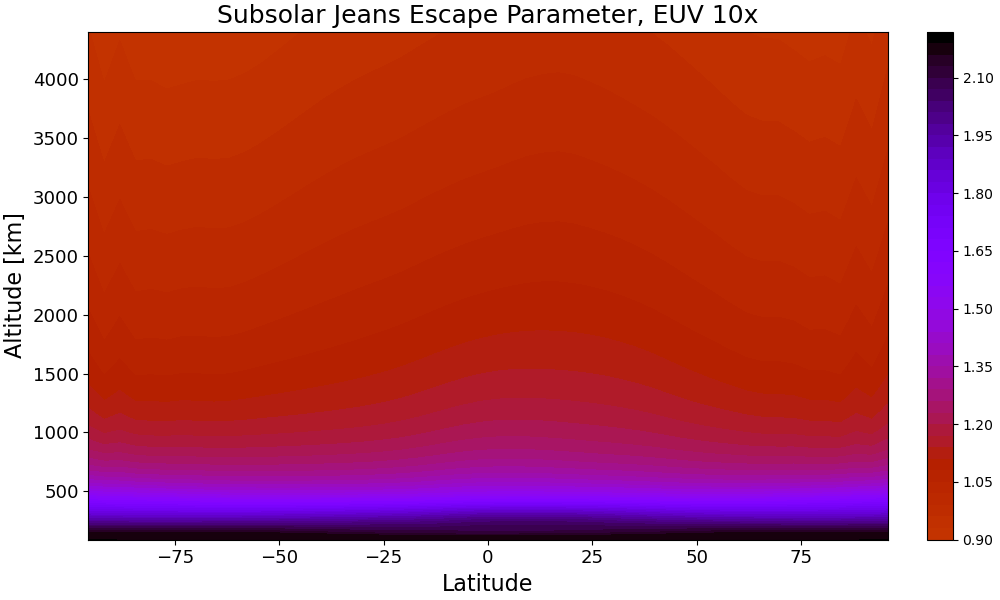}\hfill
    \includegraphics[width=0.5\linewidth]{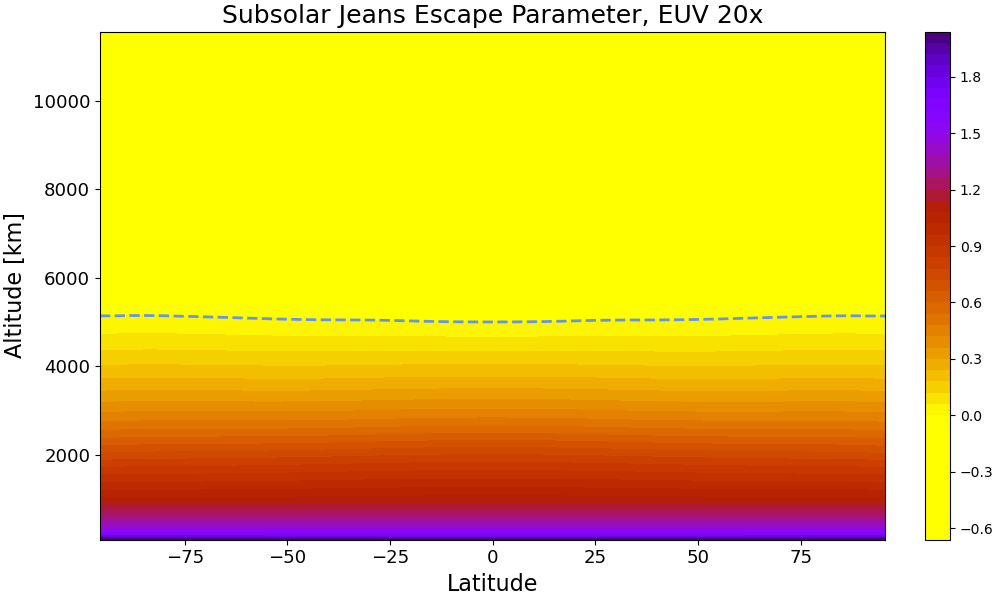}\hfill
    \includegraphics[width=0.5\linewidth]{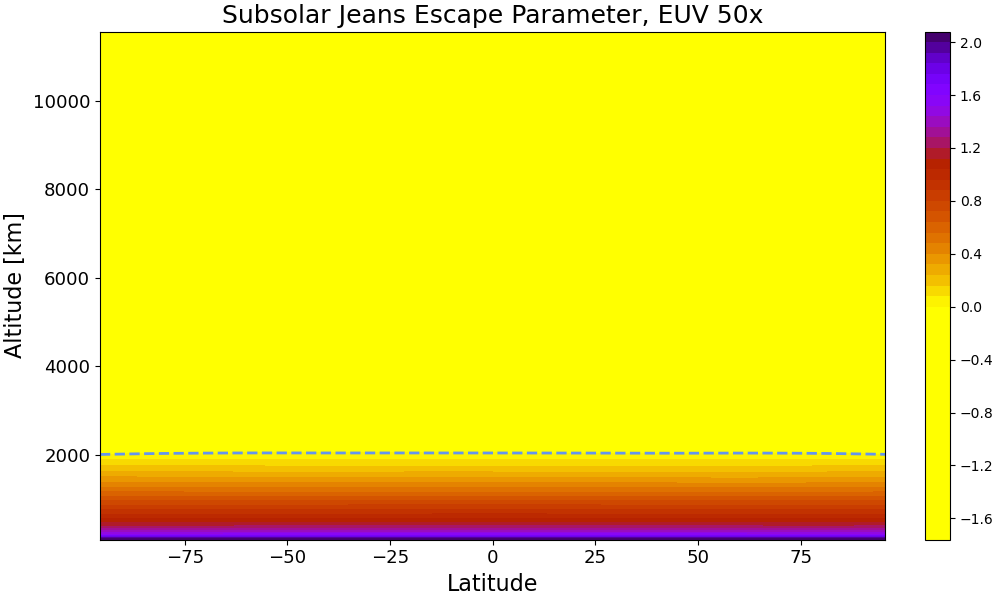}
    \caption{Log of Jean's escape parameter for $O$ calculated along along a constant longitude of 180. The dotted line follows the contour of $\log\lambda=0$, which is equivalent to $\lambda=1$ and represents the point at which particles are in complete escape.}
    \label{fig:jeansO+}
\end{figure*}

Our results show that the increased EUV radiation inflates the atmosphere and raises the altitude of the ionosphere where electron and ion densities are maximized. The peak densities of electrons and $O^+$ ions increase as more energy is deposited in the atmosphere. Our results for the $O$ cooling rates further support this by showing that EUV heating increases with the increased EUV levels but $O$ cooling remains relatively unchanged as more energy is dumped into the atmosphere. We also find that increasing the EUV radiation leads to a fairly uniform increase in thermospheric temperature for cases at or below 10 times solar levels. 

The order of magnitude increase in the vertical flux for EUV factors of 7 and 9 demonstrates that an Earth-like exoplanet subject to higher EUV levels will experience a significant increase in loss of $O^+$. In our solar system, these levels correspond to a planet that would be 2.6 -- 3 times closer to the Sun than Earth presently is. For M dwarf exoplanets, the planet would need to be much closer to experience similar levels of EUV. In order to study further the impact of the more extreme EUV cases, we calculate the Jeans escape parameter for each EUV factor, including the 20 and 50 EUV factors. The Jeans escape parameter is determined by 

\begin{equation}
    \lambda = \frac{GMm}{rkT},
\end{equation}
where $G$ is the gravitational constant, $M$ is the mass of Earth, $m$ is the mass of $O^+$, $r$ is the distance from Earth's center, $k$ is the Boltzmann constant, and $T$ is the temperature. This parameter represents the ratio of the gravitational energy of a particle of a given species at a given altitude and the average kinetic energy of particles of that species. For values above one, most of the particles will be bounded by the planetary gravitational field and only particles at the high end of the distribution function can escape (Jeans escape). For values below one, the gas thermal pressure would allow the particle population to collectively escape (Hydrodynamic escape). We have taken the logarithm of this parameter in order to more easily visualize the difference between the different EUV scale factors, so regions that are below 0 are considered to be in total escape.  

\begin{figure}
    \centering
    \includegraphics[width=\linewidth]{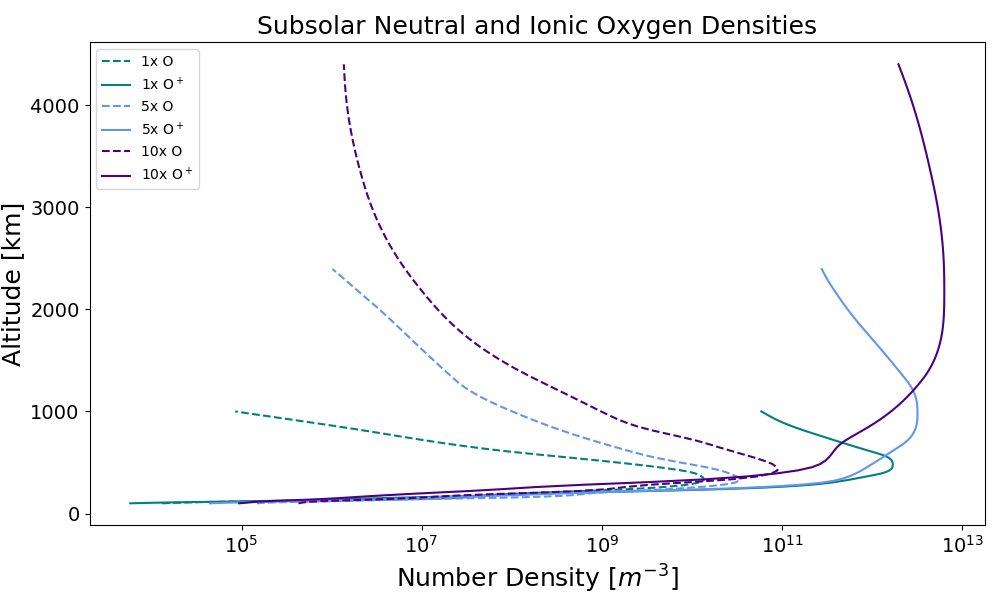}
    \caption{Comparison of subsolar neutral and ionic $O$. Neutral densities are shown with the dashed line, ion densities with the solid line. In each case shown, the ion density exceeds the neutral density by several orders of magnitude. For visual clarity we show a select number of cases, however this trend is shared with the other EUV levels not depicted.}
    \label{fig:neutralionO}
\end{figure}

Figure \ref{fig:jeansO+} shows the log of the escape parameter as a function of latitude and altitude for EUV factors of 1, 5, 9, 10, 20, and 50 along the subsolar longitude. For the EUV scale factors of 9 and 10 we see the log of the escape parameter be significantly closer to 0 than that of the scale factors of 1 and 5. When examining the even more extreme EUV levels of 20 and 50, it is immediately clear that these atmospheres are blowing off. In the case of EUV factor of 20, we see that above ~$5000~km$ the atmosphere is in the hydrodynamic escape regime, and even at lower altitudes there is high probability of hydrodynamic escape. Given this trend seen in the EUV scale factors of 20 and 50, we suspect that the 9 and 10 cases would include a region at the top of the atmosphere that is in rapid hydrodynamic escape, though we were unable to produce a stable run of these cases that was tall enough to observe this. The distribution of the escape parameter for the EUV factor of 20 solution cycled between slightly lower and higher escape probabilities similarly to the cyclic behavior shown in Fig \ref{fig:extreme_O+}. Here, we show the final time after 20 days (other times are not shown). This cyclic behavior is an artifact of the lower boundary neutral $O$ source. In a real planetary atmosphere this source may not remain constant as it is gradually depleted due to ion escape, and when factoring in the high escape probability shown in our analysis, we expect the density of $O^+$ in the atmosphere of an Earth-like planet in the subject to EUV factor of 20 to rapidly decrease. As for the EUV factor of 50 solution, we observe an atmosphere that is in nearly total hydrodynamic escape, since the log of the escape parameter is below 0 everywhere above $2000~km$. The implications of this trend in the escape parameter are significant, as they demonstrate a transition in the primary escape mechanism for these higher EUV cases compared to our baseline and other lower EUV scale factors. On Earth, the pressure gradient is not sufficient to drive both neutral and ion hydrodynamic escape. The polar wind is the primary mechanism to drive minor ion escape of $O^+$ and other ion species, and it is responsible for the observed $10^{26}s^{-1}$ loss rate of $O^+$. It is typical to ignore Jeans escape for both $O^+$ and neutral $O$ because an insignificant part of distribution function has a velocity which exceeds the Earth's escape velocity. In our numerical experiment, the increased EUV input leads to significant heating of the atmosphere so that the pressure gradient becomes the dominant driver of escape for both neutrals and ions (dominating all other escape mechanisms for neutrals and ions). Thus, hydrodynamic escape becomes the dominant escape mechanism for {\it both} $O^+$ and neutral $O$ \citep[see, e.g.,][]{2020JGRA..12527639G}. Fig \ref{fig:neutralionO} compares the neutral $O$ and $O^+$ densities for selected solutions. It can be shown that the $O^+$ densities exceed those of the neutral Oxygen ($O_2$ densities drop very rapidly with height and are even lower). Thus, the hydrodynamic escape in our domain is dominated by the ions and not the neutral Oxygen. It would be interesting to investigate how a strong neutral escape applied at our lower boundary would affect the solution. However, we leave this study for a future investigation.

In this work, we focus on the escape of $O^+$. Planets in our solar system with high levels of $CO_2$ have strong $O^+$ escape, and while that does not necessarily mean that the escape is a direct indication of habitability, it is clear that understanding how these atmospheres evolve is an important step in identifying potential habitable planets. The high EUV conditions that are imposed on an Earth-like atmosphere are intended to give us a better understanding of how an Earth-like, Nitrogen-based atmosphere might respond. M-dwarf stars are believed to have increased EUV activity, so understanding the impact of these increased levels of EUV is crucial to understand habitability of exoplanets. We use a stellar EUV spectrum as simple proxy for EUV input, however future studies could use actual, observed M-dwarf spectrum.It would also be interesting to see how a different atmosphere, such as CO$_2$-based or Methane-based atmosphere may respond differently to the variations in EUV input. Some modeling work has been done already suggests that CO$_2$ is an essential component in the development on the O$_2$ and N$_2$ dominated atmosphere Earth has today (e.g. \cite{GRASSER2023118442} \& \cite{refId0}), so understanding how different atmospheric compositions respond to increased EUV radiation will be essential to knowing how habitable atmospheres evolve. We leave this to future investigations. 

\section{Conclusion} \label{sec:conc}
We present an investigation into the impact of enhanced stellar EUV flux on the characteristics of an Earth-like atmosphere in order to study potentially habitable close orbit exoplanets. We find that 1D simulations produce significantly different results compared to 3D. Thus, 1D models are limited in their ability to capture the global atmospheric dynamics, and 3D models may be necessary in some cases. The enhanced EUV flux has a dramatic impact on the structure of the atmosphere. The density structure of ions and electrons maintains a self-similar shape, but the altitude at which peak densities increases, as well as the magnitude of the maximum density. Joule heating in the poles, as well as EUV heating increases significantly with increased EUV input. However, cooling rates do not catch up with these increases, resulting in an overall increase in the temperature of the atmosphere. An examination of the Jeans escape parameter shows that the ions are more likely to escape in the higher EUV solution, and for extreme EUV solutions of EUV factors of 20 and 50, we find atmospheres that are escaping rapidly, where it looks like the atmosphere starts to become unstable around an EUV factor of 10. Our results suggest that it is unlikely that short-orbit exoplanets around M dwarf stars would be able to sustain a habitable atmosphere. Even planets subject to just 2 or 3 times the EUV levels experienced by current Earth may not be able to sustain atmospheres for long with vertical $O^+$ fluxes exceeding $10^{27} s^{-3}$. Future investigations may include driving the model using observed EUV spectrum of actual star-planet systems.

\begin{acknowledgments}

We thank an unknown referee for their useful comments and suggestions. This work was supported by NASA grant 80NSSC23K1358. The authors acknowledge the MIT SuperCloud and Lincoln Laboratory Supercomputing Center for providing HPC resources that have contributed to the research results reported within this paper.

\end{acknowledgments}

\bibliography{references}{}
\bibliographystyle{aasjournal}



\end{document}